\documentclass[aps,pre,superscriptaddress]{revtex4}
\usepackage[dvips]{graphics}
\usepackage{graphicx}
\usepackage{amsfonts}
\usepackage{amssymb}
\usepackage{amsmath}
\usepackage{subfigure}
\usepackage[abs]{overpic}
\usepackage{color}
\usepackage{subfigure}

\begin{document}

\title{Nonlinear modes and symmetries in linearly-coupled pairs of $\mathcal{%
PT}$-invariant dimers}
\author{By K. Li, P. G. Kevrekidis and B. A. Malomed
\footnote{Address for correspondence: B. A. Malomed, 
Department of Physical Electronics, School of Electrical Engineering,
Faculty of Engineering, Tel Aviv University, Tel Aviv 69978, Israel}}
\noaffiliation
%\author{P. G.\ Kevrekidis }
%\affiliation{Department of Mathematics and Statistics, University of Massachusetts,
%Amherst MA 01003-4515, USA}
%\author{B. A. Malomed}
%\affiliation{Department of Physical Electronics, School of Electrical Engineering,
%Faculty of Engineering, Tel Aviv University, Tel Aviv 69978, Israel}

\begin{abstract}
\begin{center}
\rule{\linewidth}{0.2mm}
\end{center}
The subject of the work are pairs of linearly coupled $\mathcal{PT}$%
-symmetric dimers. Two different settings are introduced, namely, \textit{%
straight-coupled dimer}s, where each gain site is linearly coupled to one
gain and one loss site, and \textit{cross-coupled dimers}, with each gain
site coupled to two lossy ones. The latter pair with equal coupling
coefficients represents a $\mathcal{PT}$-\textit{hypersymmetric} quadrimer.
We find symmetric and antisymmetric solutions in these systems, chiefly in
an analytical form, and explore the existence, stability and dynamical
behavior of such solutions by means of numerical methods. We thus identify
bifurcations occurring in the systems, including spontaneous symmetry
breaking and saddle-center bifurcations. Simulations demonstrate that
evolution of unstable branches typically leads to blowup. However, in some
cases unstable modes rearrange into stable ones.
\begin{center}
\rule{\linewidth}{0.2mm}
\end{center}
\end{abstract}

\maketitle

%\author{K. Li\thanks{Department of Mathematics and Statistics, University of Massachusetts,
%Amherst MA 01003-4515, USA}, P. G.\ Kevrekidis\footnotemark[1], and
%B. A. Malomed\thanks{Department of Physical Electronics, School of Electrical Engineering,
%Faculty of Engineering, Tel Aviv University, Tel Aviv 69978, Israel}}

%\pacs{11.30.Er; 72.10.Fk; 42.79.Gn; 11.80.Gw}

\section{Introduction}

Recently, quantum systems and their classical wave counterparts featuring
the $\mathcal{PT}$ (parity-time) symmetry, supported by the balance between
spatially separated gain and loss terms, have drawn a great deal of
attention, as reviewed in Refs. \cite{Bender_review,special-issues,review}.
In this context, the straightforward similarity between the Schr\"{o}dinger
equation in quantum mechanics and the paraxial propagation equation in
optics has made it possible to propose~\cite{Muga,PT_periodic} and
demonstrate in experiments \cite{experiment} that the $\mathcal{PT}$ \
symmetry can be implemented in terms of the optical-beam propagation in
waveguides with appropriately placed and mutually balanced gain and loss.

The optical realizations of the $\mathcal{PT}$ symmetry make it natural to
extend this concept to nonlinear settings \cite{Musslimani2008}. In
particular, solitons can be supported by the combination of the Kerr
nonlinearity and spatially periodic complex potentials, whose odd imaginary
part accounts for the balanced gain and loss, thus accounting for the $%
\mathcal{PT}$ \ symmetry. A detailed analysis demonstrates the existence of
stability regions for such $\mathcal{PT}$-symmetric solitons, both bright
\cite{Yang} and dark ones \cite{dark,vortices}, as well as for
two-dimensional vortices~\cite{vortices}. Alternatively, one-dimensional
\cite{dual,Driben,Barash,dual2} and two-dimensional \cite{Burlak} bright $%
\mathcal{PT}$-symmetric solitons, and their one-dimensional dark
counterparts \cite{dark-coupler} can be built as stable objects in dual-core
couplers, with the balanced gain and loss placed in the different cores.
Stable bright solitons were also predicted in $\mathcal{PT}$-symmetric
settings with the second-harmonic-generating (quadratic) nonlinearity \cite%
{chi2}.

Another class of nonlinear $\mathcal{PT}$-symmetric systems is represented
by a pair of discrete (delta-functional) gain and loss elements \cite%
{Stuttgart}, or the \textit{gain-loss dipole}, with the imaginary part of
the potential represented by the $\delta ^{\prime }$ function of the
coordinate \cite{Bangkok}, which are embedded into a continuous medium with
the cubic nonlinearity. The model with the the gain-loss dipole admits a
full family of exact analytical solutions for solitons pinned to the $%
\mathcal{PT}$ dipole.

Further, discrete solitons were predicted in various chains of linear \cite%
{discrete} and circular \cite{circular} coupled $\mathcal{PT}$-symmetric
elements and, more generally, in networks of coupled $\mathcal{PT}$%
-symmetric \textit{oligomers} (dimers, quadrimers, etc.)~\cite{KPZ,KPZ0,KPZ1}%
. Parallel to incorporating the usual Kerr nonlinearity into the
conservative part of the $\mathcal{PT}$ system, its gain-loss-antisymmetric
part can be made nonlinear too, by introducing mutually balanced cubic gain
and loss terms \cite{AKKZ,we}. This can be done in the simplest way in the
context of discrete systems, by embedding nonlinear cores into linear chains
\cite{we}. Effects of combined linear and nonlinear $\mathcal{PT}$~terms on
the existence and stability of optical solitons were studied too \cite%
{combined}.

Although $\mathcal{PT}$-symmetric models belong, generally speaking, to the
class of dissipative systems, the fact that they give rise to \emph{%
continuous} families of modes, which exist due to the balance between the
separated gain and loss with equal strengths, makes them similar to
conservative systems. The $\mathcal{PT}$-symmetry of the modes gets broken
with the increase of the gain-loss coefficient. Above the critical value of
this coefficient, the solution typically undergoes blowup~\cite{pelin_recent}%
, due to the onset of the imbalance between the linear gain and loss. On the
other hand, in the presence of the nonlinear $\mathcal{PT}$-balanced gain
and loss terms, the symmetry breaking of the solutions may lead to the
formation of a self-trapped asymmetric mode, rather than the blowup \cite{we}%
. However, in the latter case the modes exist as \emph{isolated attractors }%
(rather than continuous families), like in generic nonlinear dissipative
systems, i.e., the $\mathcal{PT}$ symmetry is broken in that case too.

The purpose of the present work is to introduce nonlinear systems with
double symmetries, $\mathcal{PT}$ and spatial, and explore results of the
interplay between these symmetries. The simplest example of such a setting,
which we construct and analyze here, are \textit{bi-dimers}, i.e., pairs of
two linearly coupled $\mathcal{PT}$-symmetric dimers with the on-site cubic
nonlinearity. Two types of this setting are possible, both considered below:
\textit{straight}- and \textit{cross}-coupled ones, in which, respectively,
the gain and loss elements of one dimer are coupled to their counterparts in
the parallel one, or, alternatively, the two dimers are set anti-parallel to
each other, the gain pole of one being coupled to its lossy counterpart in
the other. In earlier works, similar configurations were considered either
for equal couplings between all the sites~\cite{KPZ1}, or for special cases
(e.g., cross-coupled dimers for a special form of unequal couplings were
touched upon in Ref.~\cite{KPZ0}). None of the earlier considered bi-dimer
settings included the above-mentioned nonlinear $\mathcal{PT}$-symmetric
gain/loss terms, which are a part of the models introduced in the present
work.

The paper is organized as follows. In Section II, we formulate the\ models
of the straight- and cross-coupled bi-dimers. We report partial analytical
and systematic numerical results, varying the linear gain/loss parameter, $%
\gamma _{0}$, in the presence of the nonlinear gain and loss terms with
coefficient $\gamma _{2}$, for varieties of stationary modes in the
straight- and cross-coupled bi-dimers in Sections III and IV, respectively.
In particular, the cross-coupled bi-dimer with the two linear-coupling
constants equal to each other may be considered as a $\mathcal{PT}$\textit{%
-hypersymmetric} quadrimer. The paper is concluded by Section V, which also
puts forward directions for future studies.

\section{Formulation of the models}

\subsection{The straight-coupled bi-dimer}

\begin{figure}[tph]
\label{config} \scalebox{0.5}{\includegraphics{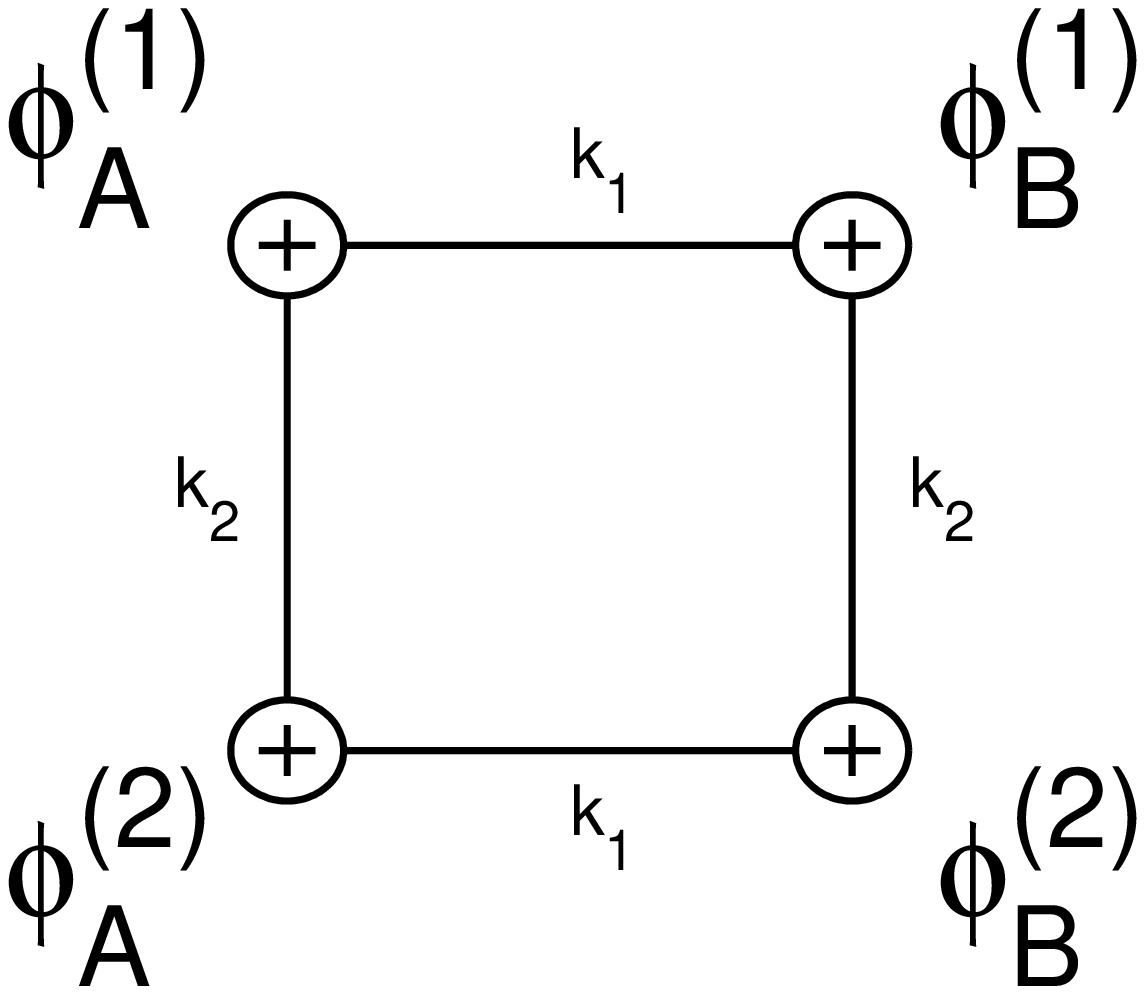}} %
\scalebox{0.5}{\includegraphics{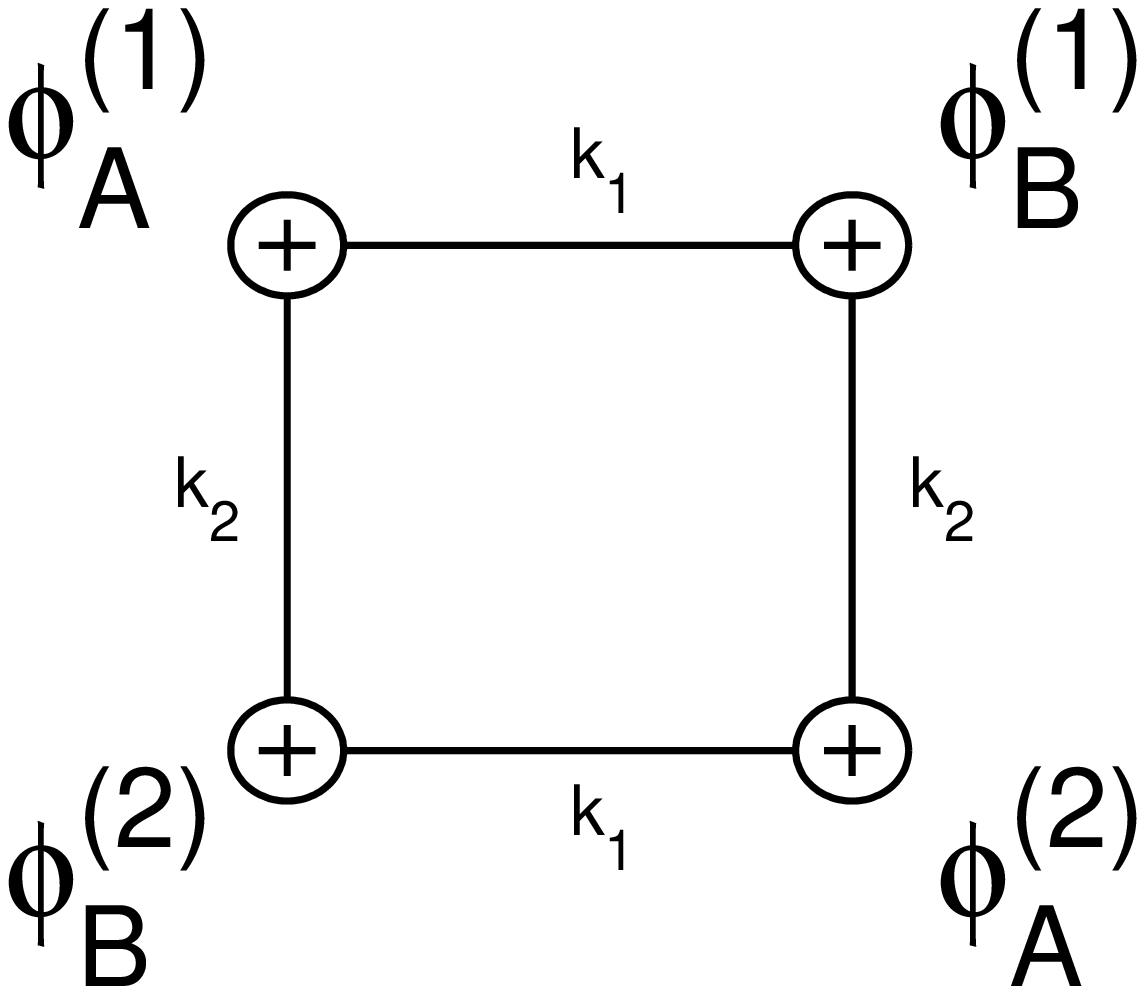}}
\caption{The left and right panels schematically display the straight- and
cross-coupled bi-dimers, respectively.}
\label{straightcross}
\end{figure}

We introduce a system of two linearly coupled $\mathcal{PT}$-symmetric
dipoles, each one represented by complex variables $\psi _{A,B}^{(1,2)}$,
where subscripts $A$ and $B$ stand for the gain and loss sites, while
superscripts $1$ and $2$ refer to the dipole's number. Stationary states
with frequency $\omega $ are looked for in the form of
\begin{equation}
~\psi _{A,B}^{(1,2)}(t)=e^{-i\omega t}\phi _{A,B}^{(1,2)}.  \label{phi}
\end{equation}%
The bi-dimers with the straight and cross couplings between the gain and
loss sites are schematically shown, in terms of variables $\phi
_{A,B}^{(1,2)}$, in the left and right panels of Fig. \ref{straightcross}.

Taking the dimer elements as defined in Ref. \cite{we} (which includes the
conservative cubic nonlinearity with real coefficient $\chi $), the
straight-coupled bi-dimer is described by the following dynamical equations:
\begin{equation}
\begin{array}{l}
{i\dot{\psi}_{A}^{(1)}=}\left( {i\gamma _{0}-i\gamma _{2}}\left\vert {\psi
_{A}^{(1)}}\right\vert {^{2}+\chi }\left\vert {\psi _{A}^{(1)}}\right\vert {%
^{2}}\right) {\psi _{A}^{(1)}+k_{1}\psi _{B}^{(1)}+k_{2}\psi }_{A}^{(2)}{,}
\\*[9pt]
{i\dot{\psi}_{B}^{(1)}=}\left( {-i\gamma _{0}+i\gamma _{2}}\left\vert {\psi
_{B}^{(1)}}\right\vert {^{2}+\chi }\left\vert {\psi _{B}^{(1)}}\right\vert {%
^{2}}\right) {\psi _{B}^{(1)}+k_{1}\psi _{A}^{(1)}{+k_{2}\psi }_{B}^{(2)},}%
\end{array}
\label{v1}
\end{equation}%
\begin{equation}
\begin{array}{l}
{i\dot{\psi}_{A}^{(2)}=}\left( {i\gamma _{0}-i\gamma _{2}}\left\vert {\psi
_{A}^{(2)}}\right\vert {^{2}+\chi }\left\vert {\psi _{A}^{(2)}}\right\vert {%
^{2}}\right) {\psi _{A}^{(2)}+k_{1}\psi _{B}^{(2)}+k_{2}\psi }_{A}^{(1)}{,}
\\*[9pt]
{i\dot{\psi}_{B}^{(2)}=}\left( {-i\gamma _{0}+i\gamma _{2}}\left\vert {\psi
_{B}^{(2)}}\right\vert {^{2}+\chi }\left\vert {\psi _{B}^{(2)}}\right\vert {%
^{2}}\right) {\psi _{B}^{(2)}+k_{1}\psi _{A}^{(2)}{+k_{2}\psi }_{B}^{(1)},}%
\end{array}
\label{v2}
\end{equation}%
where $\gamma _{0}>0$ and $\gamma _{2}>0$ are the linear and nonlinear
gain-loss coefficient, $k_{1}$ accounts for the linear coupling, inside a
give dimer, between the sites at which the gain and loss are applied, and $%
k_{2}$ is a coefficient of the coupling between the parallel dimers. Then,
stationary solutions are looked as per Eq. (\ref{phi}) with constant
amplitudes $\phi _{A,B}^{(1,2)}$ satisfying a system of algebraic equations:%
\begin{equation}
\begin{array}{l}
{\omega \phi _{A}^{(1)}=}\left( {i\gamma _{0}-i\gamma _{2}}\left\vert {\phi
_{A}^{(1)}}\right\vert {^{2}+\chi }\left\vert {\phi _{A}^{(1)}}\right\vert {%
^{2}}\right) {\phi _{A}^{(1)}+k_{1}\phi _{B}^{(1)}+k_{2}{\phi }_{A}^{(2)},}
\\*[9pt]
{\omega \phi _{B}^{(1)}=}\left( {-i\gamma _{0}+i\gamma _{2}}\left\vert {\phi
_{B}^{(1)}}\right\vert {^{2}+\chi }\left\vert {\phi _{B}^{(1)}}\right\vert {%
^{2}}\right) {\phi _{B}^{(1)}+k_{1}\phi _{A}^{(1)}{+k_{2}\phi }_{B}^{(2)},}%
\end{array}
\label{phi1}
\end{equation}%
\begin{equation}
\begin{array}{l}
{\omega \phi _{A}^{(2)}=}\left( {i\gamma _{0}-i\gamma _{2}}\left\vert {\phi
_{A}^{(2)}}\right\vert {^{2}+\chi }\left\vert {\phi _{A}^{(2)}}\right\vert {%
^{2}}\right) {\phi _{A}^{(2)}+k_{1}\phi _{B}^{(2)}+k_{2}{\phi }_{A}^{(1)},}
\\*[9pt]
{\omega \phi _{B}^{(2)}=}\left( {-i\gamma _{0}+i\gamma _{2}}\left\vert {\phi
_{B}^{(2)}}\right\vert {^{2}+\chi }\left\vert {\phi _{B}^{(2)}}\right\vert {%
^{2}}\right) {\phi _{B}^{(2)}+k_{1}\phi _{A}^{(2)}{+k_{2}\phi }_{B}^{(1)}.}%
\end{array}
\label{phi2}
\end{equation}%
Notice that in this model each site with linear gain features nonlinear loss
and vice versa, as such a setting is likely to produce stable states \cite%
{we}.

\subsection{The cross-coupled bi-dimer}

The system of two antiparallel linearly (cross-) coupled dimers is described
by the following dynamical and static equations, cf. Eqs. (\ref{v1})-(\ref%
{phi2}):%
\begin{equation}
\begin{array}{l}
{i\dot{\psi}_{A}^{(1)}=}\left( {i\gamma _{0}-i\gamma _{2}}\left\vert {\psi
_{A}^{(1)}}\right\vert {^{2}+\chi }\left\vert {\psi _{A}^{(1)}}\right\vert {%
^{2}}\right) {\psi _{A}^{(1)}+k_{1}\psi _{B}^{(1)}+k_{2}\psi }_{B}^{(2)}{,}
\\*[9pt]
{i\dot{\psi}_{B}^{(1)}=}\left( {-i\gamma _{0}+i\gamma _{2}}\left\vert {\psi
_{B}^{(1)}}\right\vert {^{2}+\chi }\left\vert {\psi _{B}^{(1)}}\right\vert {%
^{2}}\right) {\psi _{B}^{(1)}+k_{1}\psi _{A}^{(1)}{+k_{2}\psi }_{A}^{(2)},}%
\end{array}
\label{q1}
\end{equation}%
\begin{equation}
\begin{array}{l}
{i\dot{\psi}_{A}^{(2)}=}\left( {i\gamma _{0}-i\gamma _{2}}\left\vert {\psi
_{A}^{(2)}}\right\vert {^{2}+\chi }\left\vert {\psi _{A}^{(2)}}\right\vert {%
^{2}}\right) {\psi _{A}^{(2)}+k_{1}\psi _{B}^{(2)}+k_{2}\psi }_{B}^{(1)}{,}
\\*[9pt]
{i\dot{\psi}_{B}^{(2)}=}\left( {-i\gamma _{0}+i\gamma _{2}}\left\vert {\psi
_{B}^{(2)}}\right\vert {^{2}+\chi }\left\vert {\psi _{B}^{(2)}}\right\vert {%
^{2}}\right) {\psi _{B}^{(2)}+k_{1}\psi _{A}^{(2)}{+k_{2}\psi }_{A}^{(1)},}%
\end{array}
\label{q2}
\end{equation}%
\begin{equation}
\begin{array}{l}
{\omega \phi _{A}^{(1)}=}\left( {i\gamma _{0}-i\gamma _{2}}\left\vert {\phi
_{A}^{(1)}}\right\vert {^{2}+\chi }\left\vert {\phi _{A}^{(1)}}\right\vert {%
^{2}}\right) {\phi _{A}^{(1)}+k_{1}\phi _{B}^{(1)}+k_{2}{\phi }_{B}^{(2)},}
\\*[9pt]
{\omega \phi _{B}^{(1)}=}\left( {-i\gamma _{0}+i\gamma _{2}}\left\vert {\phi
_{B}^{(1)}}\right\vert {^{2}+\chi }\left\vert {\phi _{B}^{(1)}}\right\vert {%
^{2}}\right) {\phi _{B}^{(1)}+k_{1}\phi _{A}^{(1)}{+k_{2}\phi }_{A}^{(2)},}%
\end{array}
\label{q1stat}
\end{equation}%
\begin{equation}
\begin{array}{l}
{\omega \phi _{A}^{(2)}=}\left( {i\gamma _{0}-i\gamma _{2}}\left\vert {\phi
_{A}^{(2)}}\right\vert {^{2}+\chi }\left\vert {\phi _{A}^{(2)}}\right\vert {%
^{2}}\right) {\phi _{A}^{(2)}+k_{1}\phi _{B}^{(2)}+k_{2}{\phi }_{B}^{(1)},}
\\*[9pt]
{\omega \phi _{B}^{(2)}=}\left( {-i\gamma _{0}+i\gamma _{2}}\left\vert {\phi
_{B}^{(2)}}\right\vert {^{2}+\chi }\left\vert {\phi _{B}^{(2)}}\right\vert {%
^{2}}\right) {\phi _{B}^{(2)}+k_{1}\phi _{A}^{(2)}{+k_{2}\phi }_{A}^{(1)}.}%
\end{array}
\label{q2stat}
\end{equation}%
Here $k_{2}$ is again a real coupling constant. As said above, in the
cross-coupled bi-dimers, each site with a linear gain is coupled to two
sites with linear loss (and vice-versa). In particular, the case of the $%
\mathcal{PT}$ \textit{hypersymmetry} (alias double symmetry) corresponds to $%
k_{2}=k_{1}$ (in the continuous model of the $\mathcal{PT}$-symmetric
coupler, the extended symmetry of the same type was introduced in Refs. \cite%
{dual} and \cite{Driben}, under the name of ``supersymmetry", which
we do not use here, to avoid confusion with the well-known
supersymmetry between bosons and fermions in the quantum field
theory). In the latter case, the cross-coupled bi-dimer may also be
naturally called a $\mathcal{PT}$-\textit{hypersymmetric
quadrupole}.

In the particular case of $\chi =0$ and $\omega =0$, solutions to the
hypersymmetric version of Eqs. (\ref{q1stat}) and (\ref{q2stat}) may be
sought for in the form similar to that in the case of the single dimer \cite%
{we}, \textit{viz}.,
\begin{equation}
\phi _{A}^{(1,2)}=A_{1,2},~\phi _{B}^{(1,2)}=iB_{1,2},  \label{simple2}
\end{equation}%
where real amplitudes $A_{1,2}$ and $B_{1,2}$ obey the following system of
four equations:%
\begin{eqnarray}
\left( \gamma _{0}-\gamma _{2}A_{1}^{2}\right) A_{1} &=&\left( \gamma
_{0}-\gamma _{2}A_{2}^{2}\right) A_{2}=-k_{1}\left( B_{1}+B_{2}\right) ,
\notag \\
\left( \gamma _{0}-\gamma _{2}B_{1}^{2}\right) B_{1} &=&\left( \gamma
_{0}-\gamma _{2}B_{2}^{2}\right) B_{2}=-k_{1}\left( A_{1}+A_{2}\right) .
\label{AB}
\end{eqnarray}%
In particular, a corollary of Eqs. (\ref{AB}) is the following relations
between the amplitudes:%
\begin{equation}
A_{1}^{2}+A_{2}^{2}+A_{1}A_{2}=B_{1}^{2}+B_{2}^{2}+B_{1}B_{2}=\gamma
_{0}/\gamma _{2}.
\end{equation}%
Below, we address the existence, stability and dynamics of nonlinear modes
in both the straight- and cross-coupled bi-dimers.

\section{Straight-coupled bi-dimers}

In this section, we first analytically seek for stationary solutions with
(real) frequency $\omega $, as per Eqs.~(\ref{phi1})-(\ref{phi2}). Then, we
will numerically explore the linear stability and nonlinear dynamics of
these solutions.

\subsection{Solutions for stationary modes}

Using the amplitude-phase parametrization for the complex variables,%
\begin{equation}
\phi _{A}^{(1)}=Ae^{i\theta _{1}},\ \phi _{B}^{(1)}=Be^{i\theta _{2}},\ \phi
_{A}^{(2)}=Ce^{i\theta _{3}},\phi _{B}^{(2)}=De^{i\theta _{4}},  \label{ABCD}
\end{equation}%
we have found nine branches of solutions to stationary equations~(\ref{phi1}%
) and (\ref{phi2}), in both analytical and numerical forms, which are listed
below.

\begin{enumerate}
\item Two solutions, which correspond to signs $\pm $ in the expression for $%
A^{2}$  in Eq. (\ref{eqstraight12}), with the unbroken spatial \textit{%
antisymmetry}~\footnote{%
When we refer to a symmetry as unbroken, we mean that it is
preserved; i.e., in this case, the solution is anti-symmetric, as is
made evident by its explicit form. Using term ``spatial", we refer
to how the waveform with superscript (2) relates to the one with
superscript (1). Thus, the spatial antisymmetry and symmetry imply,
respectively, $\phi _{A,B}^{(1)}=-\ \phi _{A,B}^{(2)}$ and $\phi
_{A,B}^{(1)}=\phi _{A,B}^{(2)}$. On the other hand, the
$\mathcal{PT}$ symmetry or its breaking refer to the field
amplitudes with the same spatial superscript, showing how the
amplitudes with subscript $A$ relate to ones with subscript $B$.
Thus, the $\mathcal{PT}$ symmetry takes place when the
amplitudes at the gain and loss sites have equal absolute values: $%
\left\vert \phi _{A}^{(1,2)}\right\vert ^{2}=\left\vert \phi
_{B}^{(1,2)}\right\vert ^{2}$, while it is broken when they are unequal.}
and unbroken $\mathcal{PT}$ symmetry: $\phi _{A}^{(1)}=-\ \phi _{A}^{(2)}$, $%
\phi _{B}^{(1)}=-\phi _{B}^{(2)}$, $\left\vert \phi _{A}^{(1,2)}\right\vert
^{2}=\left\vert \phi _{B}^{(1,2)}\right\vert ^{2}$:
\begin{gather}
A=B=C=D,  \notag \\
\theta _{1}-\theta _{3}=\theta _{2}-\theta _{4}=\pi ,  \notag \\
A^{2}=\frac{\gamma _{0}\gamma _{2}+\chi (k_{2}+\omega )\pm \sqrt{%
k_{1}^{2}(\gamma _{2}^{2}+\chi ^{2})-(\gamma _{0}\chi -\gamma
_{2}(k_{2}+\omega ))^{2}}}{\gamma _{2}^{2}+\chi ^{2}}~,  \notag \\
\sin (\theta _{1}-\theta _{2})=\frac{\gamma _{0}-\gamma _{2}A^{2}}{k_{1}},
\notag \\
\cos (\theta _{1}-\theta _{2})=\frac{k_{2}+\omega -\chi A^{2}}{k_{1}},
\label{eqstraight12}
\end{gather}%
Note that this solution satisfies the self-consistency condition, $\sin
^{2}(\theta _{1}-\theta _{2})+\cos ^{2}(\theta _{1}-\theta _{2})\equiv 1$.
Of course, here and below only the solutions with $A^{2}>0$ are meaningful
ones.

\item Two solutions with unbroken spatial symmetry and broken $\mathcal{PT}$
symmetry: $\phi _{A}^{(1)}=\ \phi _{A}^{(2)}$, $\phi _{B}^{(1)}=\phi
_{B}^{(2)}$, $\left\vert \phi _{A}^{(1,2)}\right\vert ^{2}\neq \left\vert
\phi _{B}^{(1,2)}\right\vert ^{2}$. These solution branches are non-generic
(of codimension $1$), existing under a special condition,
\begin{equation}
\chi \gamma _{0}=(k_{2}-\omega )\gamma _{2}.  \label{special}
\end{equation}%
If this condition holds, the two analytical solutions are%
\begin{gather}
A=C,\ B=D,\ A^{2}+B^{2}=\frac{\gamma _{0}}{\gamma _{2}},  \notag \\
A^{2}=\frac{\gamma _{0}\left( \gamma _{2}^{2}+\chi ^{2}\right) \pm \sqrt{%
\left( \gamma _{2}^{2}+\chi ^{2}\right) \left( -4k_{1}^{2}\gamma
_{2}^{2}+\gamma _{0}^{2}\left( \gamma _{2}^{2}+\chi ^{2}\right) \right) }}{%
2\gamma _{2}\left( \gamma _{2}^{2}+\chi ^{2}\right) },  \notag \\
\sin (\theta _{1}-\theta _{2})=\frac{(\gamma _{0}-\gamma _{2}A^{2})A}{k_{1}B}%
,  \notag \\
\cos (\theta _{1}-\theta _{2})=\frac{(\omega -k_{2}-\chi A^{2})A}{k_{1}B}.
\label{eqstraight45}
\end{gather}

Further, the point of the spontaneous breakup of the spatial symmetry of
solution (\ref{eqstraight45}), which should lead to \textit{fully asymmetric}
modes, can be found by looking for perturbed stationary solutions, $\phi
_{A}^{(1,2)}=A\pm \delta \phi _{A}$, $\phi _{B}^{(1,2)}=iB\pm \delta \phi
_{B}$, with infinitesimally small $\delta \phi _{A,B}$ . The substitution of
this into Eqs. (\ref{phi1}), (\ref{phi2}) and the linearization with respect
to the perturbations leads to a system of linear homogeneous equations,%
\begin{eqnarray}
\left( 2k_{2}-i\gamma _{0}+2i\gamma _{2}A^{2}\right) \delta \phi
_{A}+i\gamma _{2}A^{2}\delta \phi _{A}^{\ast }-k_{1}\delta \phi _{B} &=&0,
\notag \\
\left( 2k_{2}+i\gamma _{0}-2i\gamma _{2}B^{2}\right) \delta \phi
_{A}+i\gamma _{2}B^{2}\delta \phi _{B}^{\ast }-k_{1}\delta \phi _{A} &=&0.
\label{linear}
\end{eqnarray}%
Splitting the complex perturbations into real and imaginary parts, $\delta
\phi _{A,B}\equiv \delta \phi _{A,B}^{\prime }+i\delta \phi _{A,B}^{\prime
\prime },$ transforms Eqs. (\ref{linear}) into a system of four equations,
whose solvability condition yields an equation which determines the
aforementioned point of the spontaneous symmetry breaking:%
\begin{equation}
\left\vert
\begin{array}{cccc}
2k_{2} & \gamma _{0}-\gamma _{2}A^{2} & -k_{1} & 0 \\
-\gamma _{0}+3\gamma _{2}A^{2} & 2k_{2} & 0 & -k_{1} \\
-k_{1} & 0 & 2k_{2} & 2\gamma _{0}-3\gamma _{2}A^{2} \\
0 & -k_{1} & \gamma _{2}A^{2} & 2k_{2}%
\end{array}%
\right\vert =0.  \label{det}
\end{equation}%
The value of $k_{2}$ at which the symmetry breaking occurs can be found from
Eq. (\ref{det}) in an analytical form:
\begin{gather}
k_{2}=\frac{1}{8}\left( 2k_{1}^{2}-\gamma _{0}^{2}+6A^{2}\gamma _{0}\gamma
_{2}-6A^{4}\gamma _{2}^{2}\right.  \\
\left. \pm \sqrt{-12k_{1}^{2}\gamma _{0}^{2}+\gamma
_{0}^{4}+64A^{2}k_{1}^{2}\gamma _{0}\gamma _{2}-4A^{2}\gamma _{0}^{3}\gamma
_{2}-64A^{4}k_{1}^{2}\gamma _{2}^{2}+4A^{4}\gamma _{0}^{2}\gamma _{2}^{2}}%
\right) .
\end{gather}

%In particular, the possibility of the spatially asymmetric solution at %$%
%%k_2 \rightarrow 0$
%is suggested by the fact that Eq. (\ref{eqstraight45}) has two different
%physical solutions for $A^{2}$, which may represent the coexisting
%asymmetric states in the two weakly coupled dimers.

\item Two solutions with the unbroken spatial $\mathcal{PT}$ symmetries, $%
\phi _{A}^{(1)}=\ \phi _{A}^{(2)}$, $\phi _{B}^{(1)}=\phi _{B}^{(2)}$, $%
\left\vert \phi _{A}^{(1,2)}\right\vert ^{2}=\left\vert \phi
_{B}^{(1,2)}\right\vert ^{2}$:
\begin{gather}
A^{2}=B^{2}=C^{2}=D^{2}=\frac{\gamma _{0}\gamma _{2}+\chi (\omega -k_{2})\pm
\sqrt{k_{1}^{2}(\gamma _{2}^{2}+\chi ^{2})-(\gamma _{0}\chi +\gamma
_{2}(k_{2}-\omega ))^{2}}}{\gamma _{2}^{2}+\chi ^{2}},  \notag \\
\sin (\theta _{1}-\theta _{2})=\frac{\gamma _{0}-\gamma _{2}A^{2}}{k_{1}},
\notag \\
\cos (\theta _{1}-\theta _{2})=\frac{\omega -k_{2}-\chi A^{2}}{k_{1}}.
\label{eqstraight67}
\end{gather}

\item Three solution branches with the broken spatial symmetry and unbroken $%
\mathcal{PT}$ symmetry, i.e., $A=B\neq C=D$, can be found only in a
numerical form. Their profiles are shown in Fig.~\ref{straight389}.
\end{enumerate}

\subsection{Stability of the stationary modes}

Since the solution branches given by Eq.~(\ref{eqstraight45}) exist only
under condition (\ref{special}), for being able to compare them with other
solutions, we fix the coefficients as $\gamma _{2}=1,\ k_{1}=1,\
k_{2}=\omega =0.1$, and $\chi =0$, unless stated otherwise, while the main
control parameter, the linear gain-loss coefficient, $\gamma _{0}$, is
subject to variation. Note that, according to Eqs.~(\ref{v1})-(\ref{v2}), at
$\chi =0$ coefficient $\gamma _{2}$ affects solely the absolute values of
the solutions, but not their phases (actually, it does not significantly
affect the stability of the solutions either). For this reason, $\gamma
_{2}=1$ is fixed here, without the loss of generality.

\begin{figure}[tph]
\scalebox{0.4}{\includegraphics{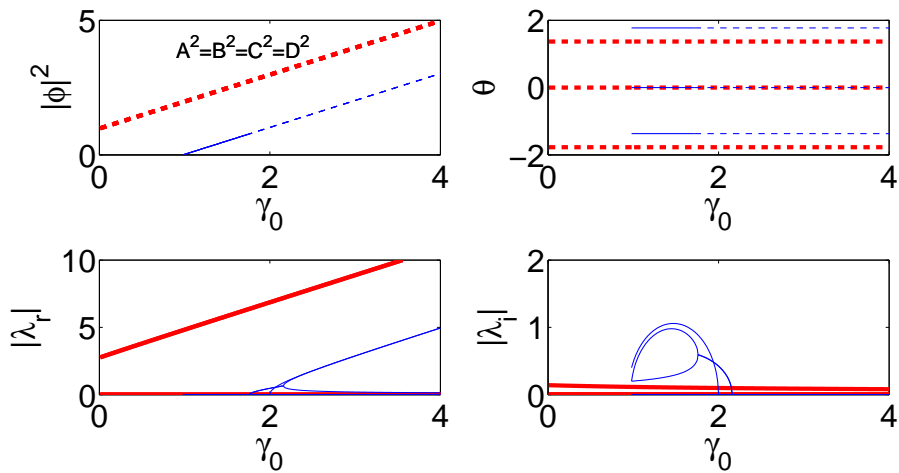}}
\caption{(Color online) Solutions for the straight-coupled bi-dimers, given
by Eq.~(\protect\ref{eqstraight12}), for fixed parameters $\protect\gamma %
_{2}=1,\ k_{2}=0.1,\ \protect\chi =0,\ k_{1}=1$, and $\protect\omega =0.1$.
Stable and unstable portions of the solution families are plotted by solid
and dashed lines, respectively. The thick red and thin blue line branches
correspond, respectively, to ``$+$" and ``$-$%
" signs in the expression for $A^{2}$ in Eq.~(\protect\ref{eqstraight12}).
The four panels show the squared absolute values (top left), phases (top
right), as well as real ($\protect\lambda _{r}$, bottom left) and imaginary (%
$\protect\lambda _{i}$, bottom right) parts of the linear stability
eigenvalues for the two branches. }
\label{straight12}
\end{figure}

\begin{figure}[tbp]
%[htp]
\scalebox{0.4}{\includegraphics{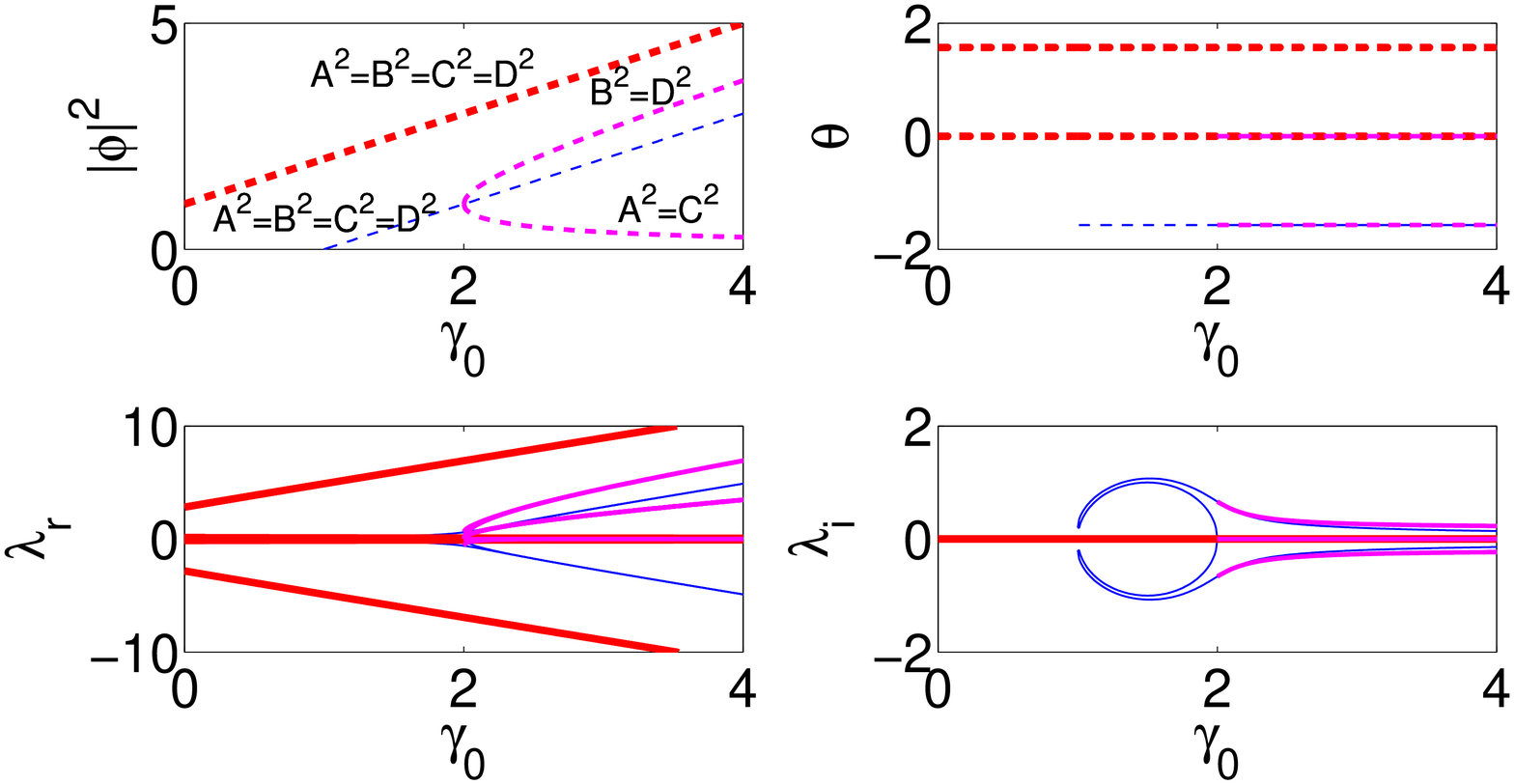}}
\caption{(Color online) Profiles of the straight-coupled bi-dimer solutions
given by Eq.~(\protect\ref{eqstraight45}) (magenta curves) and Eq.~(\protect
\ref{eqstraight67}) (thin blue and thick red ones), for fixed parameters $%
\protect\gamma _{2}=1,\ \protect\omega =k_{2}=0.1,\ \protect\chi =0$ and $%
k_{1}=1$. The thinnest blue and thickest red branches correspond to
``$-$" and ``$+$" signs in the expression for the amplitude in
Eq.~(\protect\ref{eqstraight67}), respectively. The magenta branches
with the intermediate thickness correspond to the ``$-$" sign in the
expression for $A^{2}$ in Eq.~(\protect
\ref{eqstraight45}), while the curve corresponding to the ``$%
+$" sign is plotted in Fig.~\protect\ref{straight4}.}
\label{straight4567}
\end{figure}

\begin{figure}[tbp]
%[htp]
\scalebox{0.4}{\includegraphics{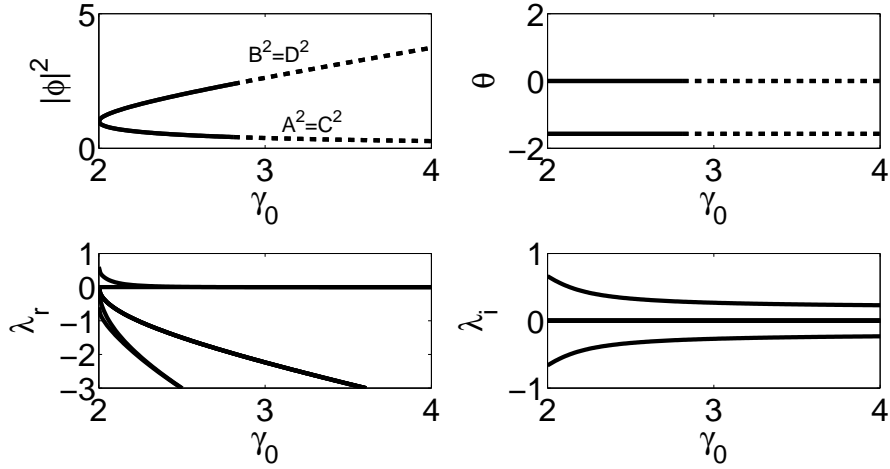}}
\caption{(Color online) Profiles of the straight-coupled bi-dimer solution
corresponding to the ``$+$" sign in Eq.~(\protect\ref%
{eqstraight45}), for fixed parameters $\protect\gamma _{2}=1,\ \protect%
\omega =k_{2}=0.1,\ \protect\chi =0$ and $k_{1}=1$.}
\label{straight4}
\end{figure}

\begin{figure}[tbp]
%[tph]
\scalebox{0.4}{\includegraphics{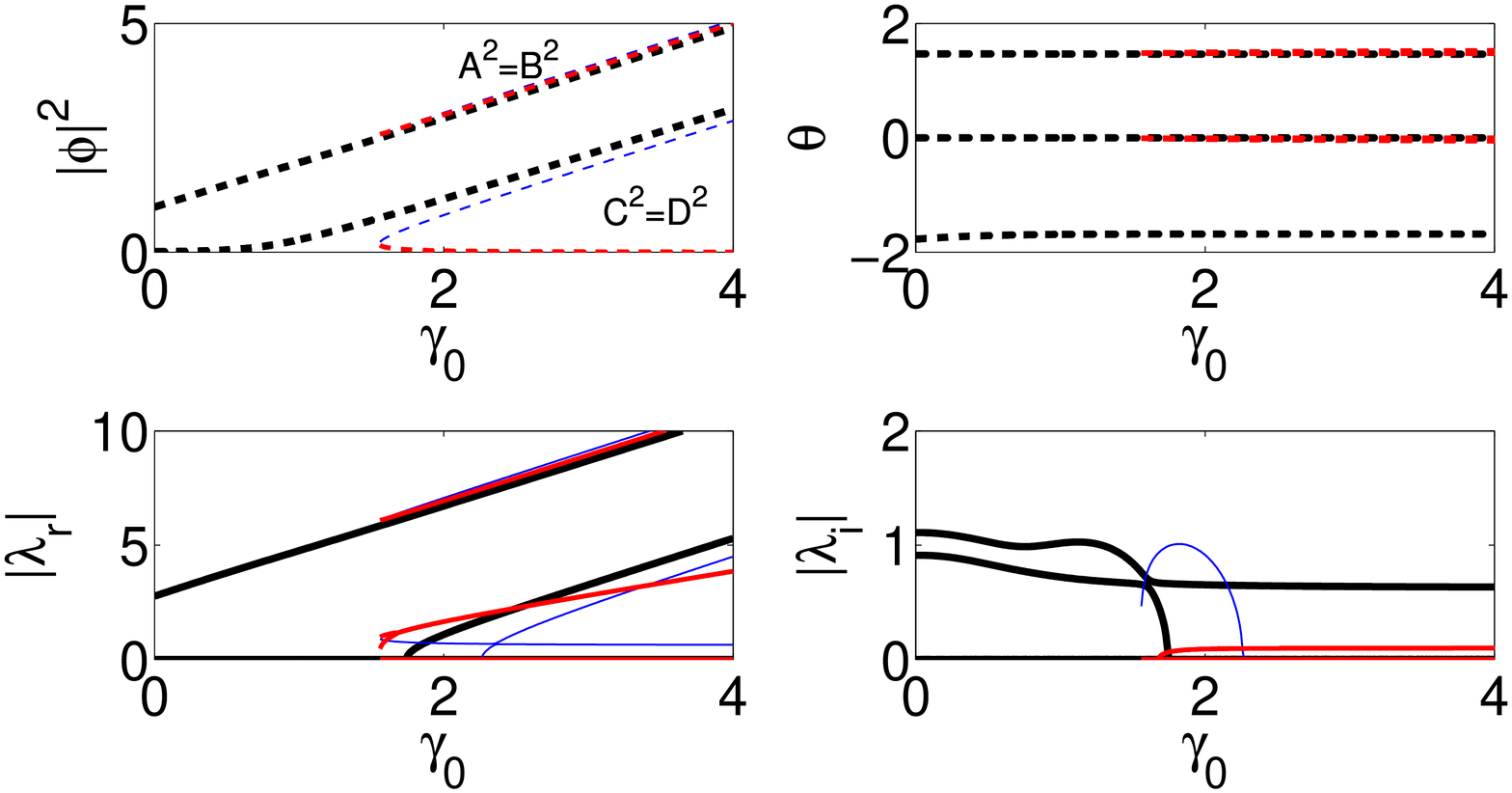}}
\caption{(Color online) Profiles of the straight-coupled bi-dimer solutions
with $A=B$ and $C=D$. Note that the phases are not constant, even if they
seem to be nearly constant, except for $\protect\theta _{1}\equiv 0$. For a
detailed description of the relevant branches see the text.}
\label{straight389}
\end{figure}

\begin{figure}[tbp]
%[tph]
\subfigure[\ thin blue
branch]{\scalebox{0.38}{\includegraphics{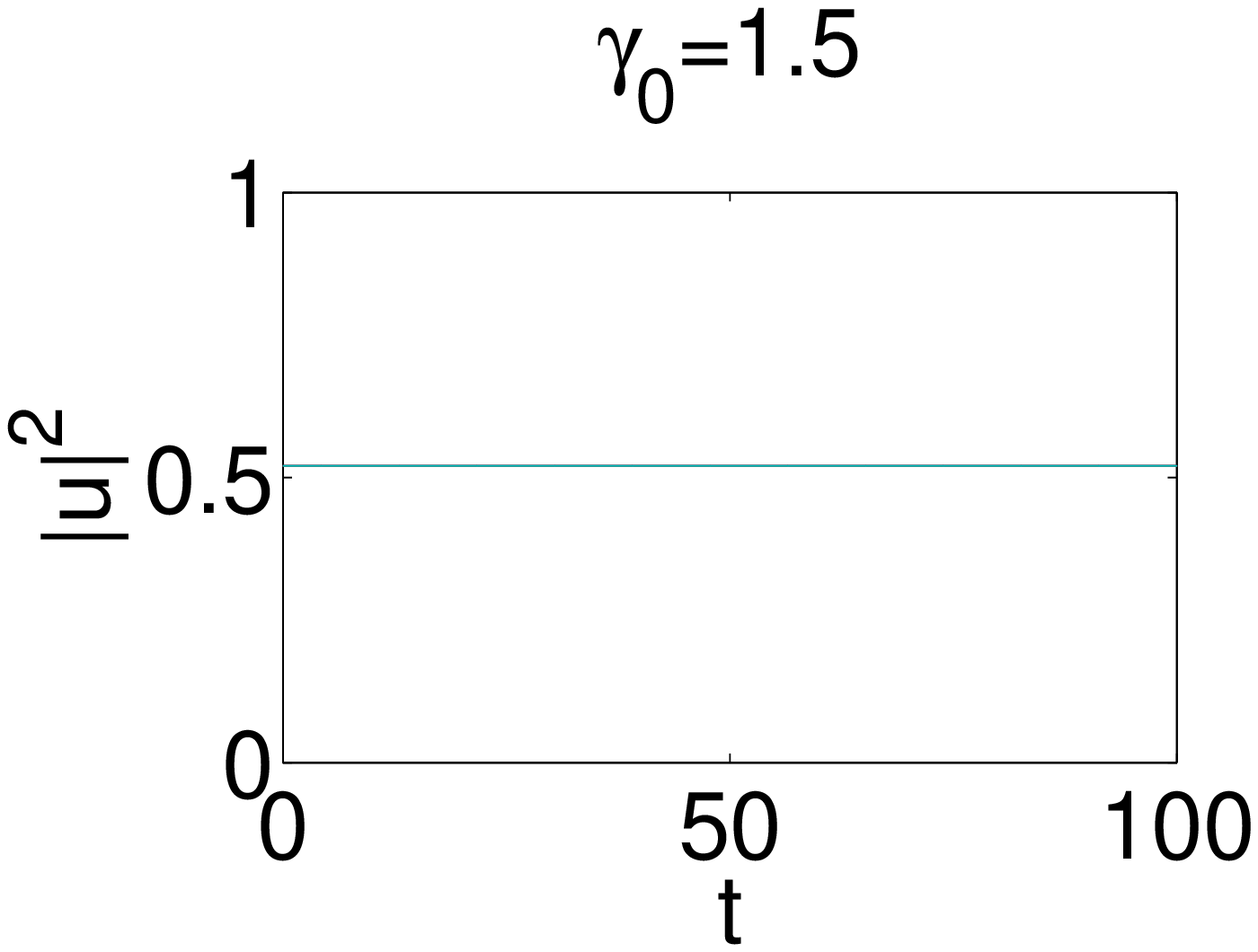}}}
\subfigure[\ thin blue
branch]{\scalebox{0.38}{\includegraphics{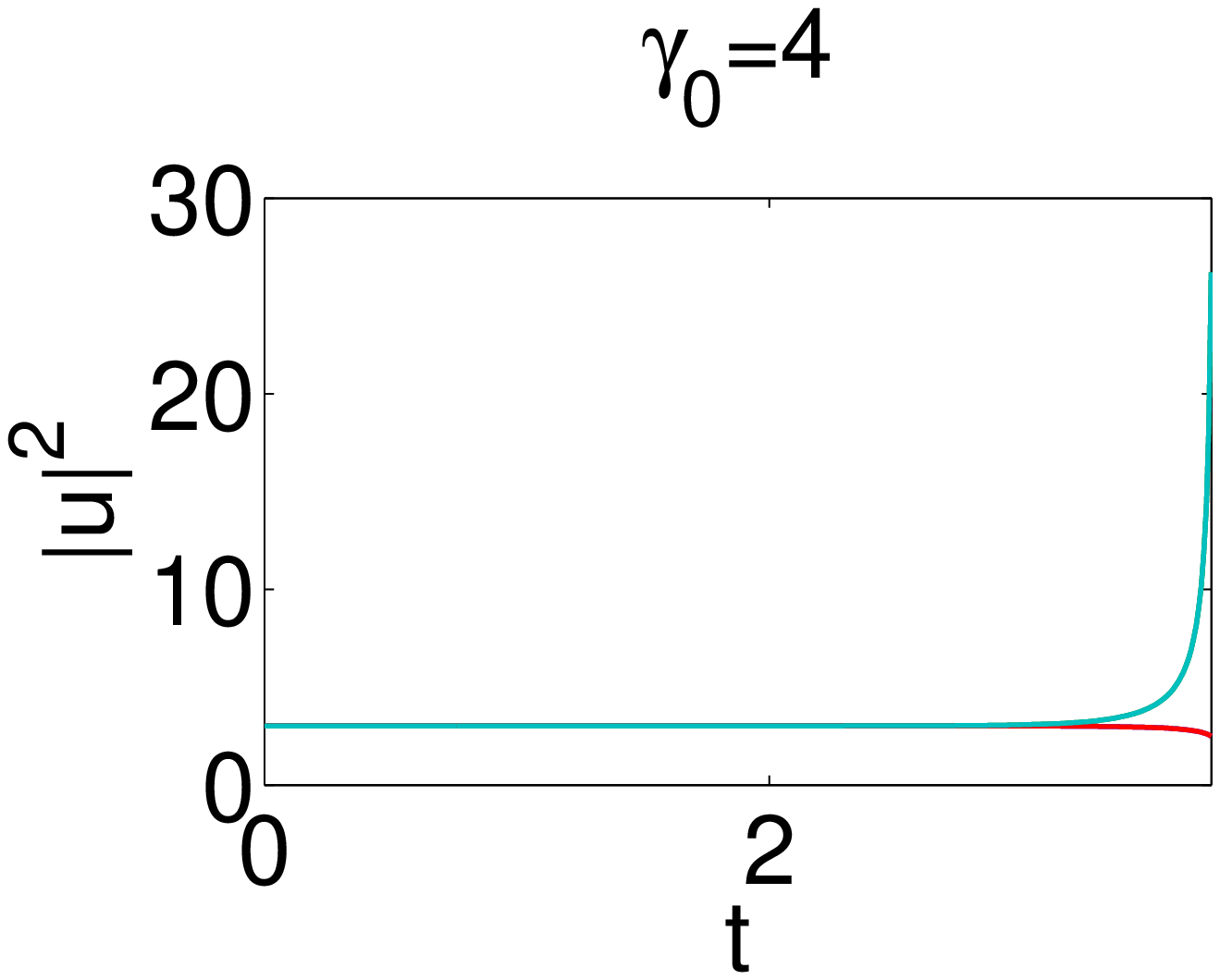}}}
\subfigure[\ thick red
branch]{\scalebox{0.38}{\includegraphics{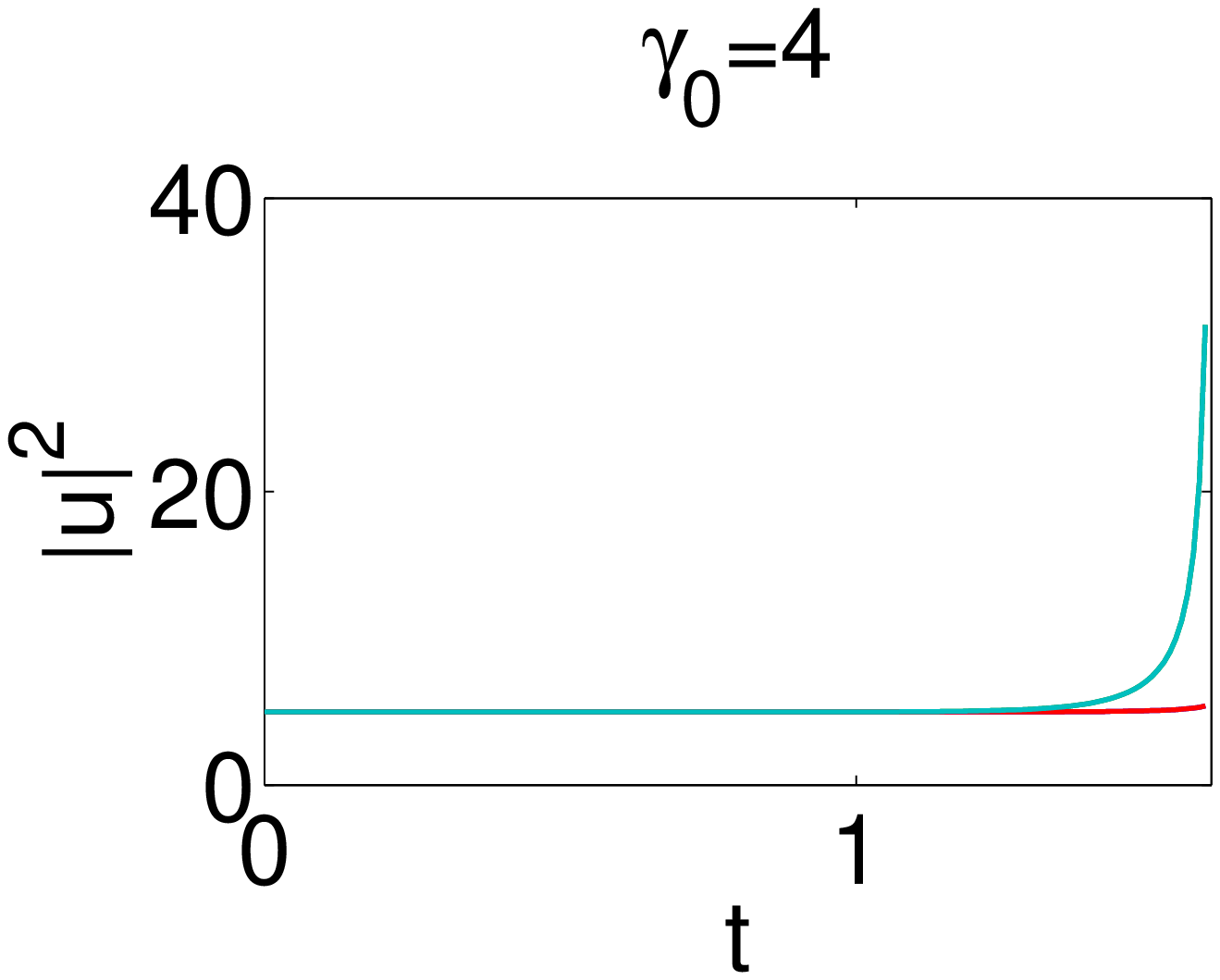}}}
% with $\gamma_0=4$. The left and right panels correspond to the blue and red branches, respectively.}
\caption{(Color online) The evolution of
the straight-coupled modes from Fig.~\protect\ref{straight12}. Here and in
dynamical simulations displayed below, the simulations were performed with
random initial perturbations of relative size $\sim 10^{-5}$ added to the
stationary solutions. Panel a) shows the stable dynamics of the solutions
represented by the thin blue branch at $\protect\gamma _{0}=1.5$. Panels b)
and c) show the instability of both thin blue and thick red branches at $%
\protect\gamma _{0}=4$, where $B\equiv D$ grows exponentially, while $%
A\equiv C$ eventually decays in b) but grows in c).}
\label{straight12_dyn}
\end{figure}

Figures~\ref{straight12}-\ref{straight12_dyn} present properties of the
straight-coupled bi-dimers of the types enumerated above. %Characteristic
%examples are displayed for parameters $\gamma _{2}=1,\ k_2 =0.1,\ \chi
%=0,\ k_1=1$ and $\omega =0.1 $, unless stated otherwise, while the main
%control parameter, the linear
%gain-loss coefficient, $\gamma _{0}$, is subject to
%variation.
The stability of the solutions was identified via numerical computation of
eigenvalues, $\lambda $, for modes of small perturbations determined by the
linearized version of Eqs.~(\ref{v1})-(\ref{v2}), which is produced by the
substitution of the expression for perturbed solutions,
\begin{equation}
\psi _{A,B}^{(1,2)}=e^{-i\omega t}\left\{ \psi _{A,B,\mathrm{eq}%
}^{(1,2)}+\delta \left[ a_{A,B}^{(1,2)}e^{\lambda t}+\left(
b_{A,B}^{(1,2)}\right) ^{\star }e^{\lambda ^{\star }t}\right] \right\} ,
\end{equation}%
into Eqs.~(\ref{v1})-(\ref{v2}), and subsequent linearization with respect
to the infinitesimal amplitude, $\delta $. The instability is implied by the
existence of a positive real part of any eigenvalue, $\lambda \equiv \lambda
_{r}+i\lambda _{i}$, with $\lambda _{r}>0$.

Figure~\ref{straight12} represents the solution branches defined by Eq.~(\ref%
{eqstraight12}), with thick red and thin blue curves corresponding to the
upper and lower signs in the expression for $A^{2}$, respectively. The thin
blue branch starts at $\gamma _{0}=0.98$, and is stable until $\gamma
_{0}=1.76$, where two pairs of purely imaginary eigenvalues collide and
create a complex quartet, thus causing the destabilization of the underlying
stationary solution through an oscillatory instability. An additional
instability arises at $\gamma _{0}=2$, through the bifurcation of an
imaginary eigenvalue pair into a real one. The thick red branch exists and
is \emph{unstable} for all values of $\gamma _{0}$. The respective
instability is accounted for by two pairs of real eigenvalues that are too
close to be distinguished in Figure~\ref{straight12}, coexisting with a pair
of purely imaginary ones.

Next, the solutions given by Eqs.~(\ref{eqstraight45}) and (\ref%
{eqstraight67}) are shown in Figs.~\ref{straight4567} and~\ref{straight4},
respectively. The black (shown in Fig.~\ref{straight4}) and magenta (the
asymmetric branch in Fig.~\ref{straight4567}) solution branches correspond,
respectively, to the upper and lower signs in the expression for $A^{2}$ in
Eq.~(\ref{eqstraight45}), while the thickest red and thinnest blue curves
represent solutions (\ref{eqstraight67}) with the upper and lower signs,
respectively. The thinnest blue branch starts at $\gamma _{0}=1$ and quickly
becomes unstable at $\gamma _{0}=1.02$ due to a pair of real eigenvalues
that arise from zero. The magenta and black branches bifurcate from the
thinnest blue one at $\gamma _{0}=2$, implying the onset of the $\mathcal{PT}
$-symmetry breaking at this point. The magenta branch is always unstable,
while its black counterpart is unstable at $\gamma _{0}<2.82$, and stable at
$\gamma _{0}>2.82$. It is important here to stress that, while for the
thinnest blue and thickest red branches $\omega $ is fixed in the course of
the continuation in $\gamma _{0}$, this is not the case for the black and
magenta branches. In particular, condition (\ref{special}) fully determines $%
\omega $ for a given set of other parameters. So, similarly to what is known
from the earlier works \cite{we,duanmu}, such non-generic branches exist at
isolated values of parameters, such as the frequency (once all other
parameters of the system are fixed).

%\textbf{PGK: Part to be commented out again}
%\textbf{I had to give up attempts to edit the following paragraph,
%as I could understand virtually nothing in it. What is shown by
%black lines? Which solutions have $A=B=C=D$, and which $A=B\neq
%C=D$? In what sense does
%the symmetry breaking occur here? If all solutions are characterized by $%
%A=B\neq C=D$, then what symmetry is broken (the spatial symmetry is already
%broken by ``}$\neq $" \textbf{in this type of the solution)?
%The blue and red curves seem to be parts of one solution, but they are
%referred to so as if they correspond to different solutions -- so, what do
%they actually represent? On the other hand, if all the solutions are
%unstable in this case, do we need to take pains of presenting this charade
%in an understandable form, or simply state that these solutions are
%completely unstable, without showing the bewildering pictures? Kai: All
%three branches shown here correspond to $A=B\neq C=D$. The blue and red are
%different branches. Their $A,B$ are very close to the $A,B$ of the black.
%(the straight line in the top left panel of Fig.~\ref{straight389}) And yes,
%they are unstable. You may consider drop them.}
%\textbf{End of part to be commented out}

We have also found three solution branches with $A=B$ and $C=D$ numerically,
which are shown in Fig.~\ref{straight389}. A bifurcation point is in this
case located at $\gamma _{0}=1.56$, when two solution branches, represented
by the thinnest blue and the thinner red lines, arise from a saddle-center
bifurcation. However, both these branches are unstable, bearing at least one
real pair of eigenvalues (the one represented by the blue line acquires a
second instability pair at $\gamma _{0}=2.26$, while the branch
corresponding to the thinner red line always possesses at least two pairs
with positive real parts). The thickest black branch exists and is unstable
for all $\gamma _{0}$ too (again, with one real pair for all values of $%
\gamma _{0}$, and an additional one emerging from the bifurcation of an
imaginary pair into a real one at $\gamma _{0}=1.74$).\textbf{\ }The
difference in the values of amplitudes $A\equiv B$ between all the three
branches is very small, but the difference in the values of $C\equiv D$
between the branches may be significant, especially for the thinner red
branch, whose $C\equiv D$ amplitudes are approaching zero as the gain/loss
parameter $\gamma _{0}$ increases. %Although it seems that
%the blue and the red branches are bifurcated from another one at $\gamma
%_{0}=1.56$, numerical result does not show the existence of such branch. But
%in principle, there could exist more branches that we did not find under the
%parameter settings.

Some case examples of the dynamical behavior of these nine branches of the
solutions are shown in Fig.~\ref{straight12_dyn}.
%,~\ref{straight4567_dyn}, and \ref{straight389_dyn}.
Among the nine above-mentioned branches, only three feature stable dynamics
in certain intervals of $\gamma _{0}$, \textit{viz}., the thin blue branch
in Fig.~\ref{straight12} for $0.98\leq \gamma _{0}\leq 1.76$, the thinnest
blue branch in Fig.~\ref{straight4567} for $1\leq \gamma _{0}\leq 1.01$, and
the black one in Fig.~\ref{straight4} for $\gamma _{0}\geq 2.82$. The
perturbed evolution of modes belonging to unstable branches always
demonstrate an indefinite growth of the amplitude (blowup) at the site where
the gain is applied, in the examples that we have considered. Amplitudes at
the loss sites decay very slowly in the solutions represented by the the
thin blue branch, whereas they grow in the solutions corresponding to the
thick red branch, though very slowly, too.

\section{Cross-coupled bi-dimers}

Similarly to the previous section, we here start by seeking for stationary
solutions with frequency $\omega $, as per Eqs. (\ref{q1stat}) and (\ref%
{q2stat}). Subsequently, we explore the linear stability and nonlinear
dynamics of such solutions.

\subsection{Solutions for stationary modes}

The cross-coupled bi-dimer complexes possess all the solutions that an
ordinary nonlinear dimer possesses (see for relevant details the recent work
\cite{duanmu}). This can be immediately seen by setting $\phi
_{A}^{(1)}=\phi _{A}^{(2)},\phi _{B}^{(1)}=\phi _{B}^{(2)}$, which reduces
Eqs. (\ref{q1stat}) and (\ref{q2stat}) to an ordinary dimer.

\begin{figure}[tbp]
%[htp]
\scalebox{0.4}{\includegraphics{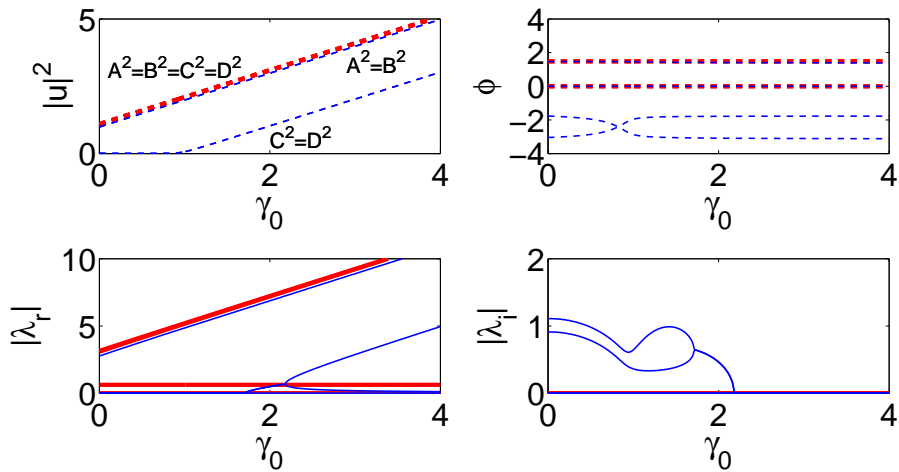}}
\caption{(Color online) Solution profiles for the cross-coupled bi-dimers at
fixed parameters $\protect\gamma _{2}=1,\ \protect\omega =k_{2}=0.1,\
\protect\chi =0$, and $\ k_{1}=1$. The thin blue and thick red lines
pertain, respectively, to the numerically found solution and the analytical
one given by Eq. (\protect\ref{cross-analyt}), respectively.}
\label{cross}
\end{figure}

Besides those obvious solutions, we have not been able to obtain other
stationary modes for cross-coupled bi-dimers in a general analytical form.
Only two branches of solutions are found in this case, without any
symmetry-breaking point, as shown in Fig.~\ref{cross}. However, it cannot be
ruled out that additional branches, potentially featuring a symmetry
breaking, may exist in this setting.

The branch with equal absolute values of the amplitudes at all four sites
(the red one in Fig. \ref{cross}), i.e., with $A=B=C=D$ (in other words, it
is a \textit{hypersymmetric quadrimer}, as it is defined above), is one that
can be found in an explicit analytical form:
\begin{gather}
A=B=C=D,  \notag \\
\phi _{1}=\phi _{3},\ \phi _{2}=\phi _{4},  \notag \\
\sin (\phi _{2}-\phi _{1})=\frac{\gamma _{2}A^{2}-\gamma _{0}}{k_{1}+k_{2}},
\notag \\
\cos (\phi _{2}-\phi _{1})=\frac{\omega -E}{k_{1}+k_{2}}.
\label{cross-analyt}
\end{gather}%
The other numerically identified solution branch is shown by the blue line
in Fig.~\ref{cross}. It is characterized by relations $A=B,\ C=D$, i.e., it
features the broken spatial symmetry and unbroken $\mathcal{PT}$ symmetry,
according to Eq. (\ref{ABCD}).

\subsection{Stability of the stationary modes}

Both solution branches (red and blue ones) for the cross-coupled bi-dimers,
which are plotted in Fig.~\ref{cross}, turn out to be unstable. The thick
red branch, which represents the analytically found solution, always has two
pairs of real eigenvalues. One of them is approximately constant, while the
other grows continuously.
%are real. One of them is fixed at $\lambda_r=\pm0.63$ while the other pair goes to infinity,
%as shown in the bottom left panel of Fig.~\ref{cross}.]}
For the thin blue branch, there is always one pair of real eigenvalues that
grows indefinitely too. On the other hand, there are two pairs of imaginary
eigenvalues that collide, giving rise to a complex quartet, and the
respective oscillatory instability, at $\gamma _{0}=1.72$. Then they collide
again at $\gamma _{0}=2.19$, turning into two pairs of real eigenvalues.

The perturbed evolution of these unstable branches is shown in Fig.~\ref%
{cross_dyn}. The evolution of the solutions corresponding to the
thin blue branch is similar to the examples discussed above for the
straightly-coupled bi-dimer, leading to the blowup (indefinite
growth). However, it is worthy to note that, unlike the other
unstable branches, which were considered above, whose amplitudes
blow up due to the instability, the evolution of solutions
associated with the thick red branch is quite different. It leads at
first to a breakup of the $\mathcal{PT}$ symmetry, bringing the pair
of amplitudes, $B=D$, very close to zero. Subsequently, the
instability results in establishment of a stable mode with squared
absolute values of the amplitudes $A^{2}=C^{2}=3.67$ and
$B^{2}=D^{2}=0.33$ (i.e., still with the broken $\mathcal{PT}$
symmetry), and phases locked to $\theta _{1}=\theta _{3},\ \theta
_{2}=\theta _{4}$ and $\theta _{1}-\theta _{2}=\pi /2$. In fact,
this eventually established solution can be identified as a stable
state of a the usual (single) nonlinear dimer, which was named
``case II" in Ref. \cite{duanmu}.
%\textbf{\ PGK: part to be
%commented out.} \textbf{[These values are very close to those for the stable
%black branch in the straight-coupled bi-dimer shown in Figs.~\ref{straight4}%
%. However, solutions with the phases locked to $0$ and $\pi /2$ are not
%admitted by stationary equations (\ref{q1stat}) and (\ref{q2stat}) for the
%cross-coupled bi-dimer, unless $k_{2}=0$. An \textit{important comment}: the
%conclusion formulated here is really perplexing. If such stationary
%solutions cannot exist, what is the meaning of the obviously stationary mode
%established in Fig. \ref{cross_dyn}(b)? Is there any stationary solution
%which corresponds to it, or it is something deeply esoteric? Kai: The red
%branch indeed converges to one of the stationary solutions of a nonlinear
%dimer. It corresponds to the case II of dimer in ~\cite{duanmu} where $%
%A^{2}+B^{2}=\gamma _{0}/\gamma _{2}$.]} End of part to be commented out.

\begin{figure}[tbp]
%[tph]
\subfigure[\ thin blue
branch]{\scalebox{0.4}{\includegraphics{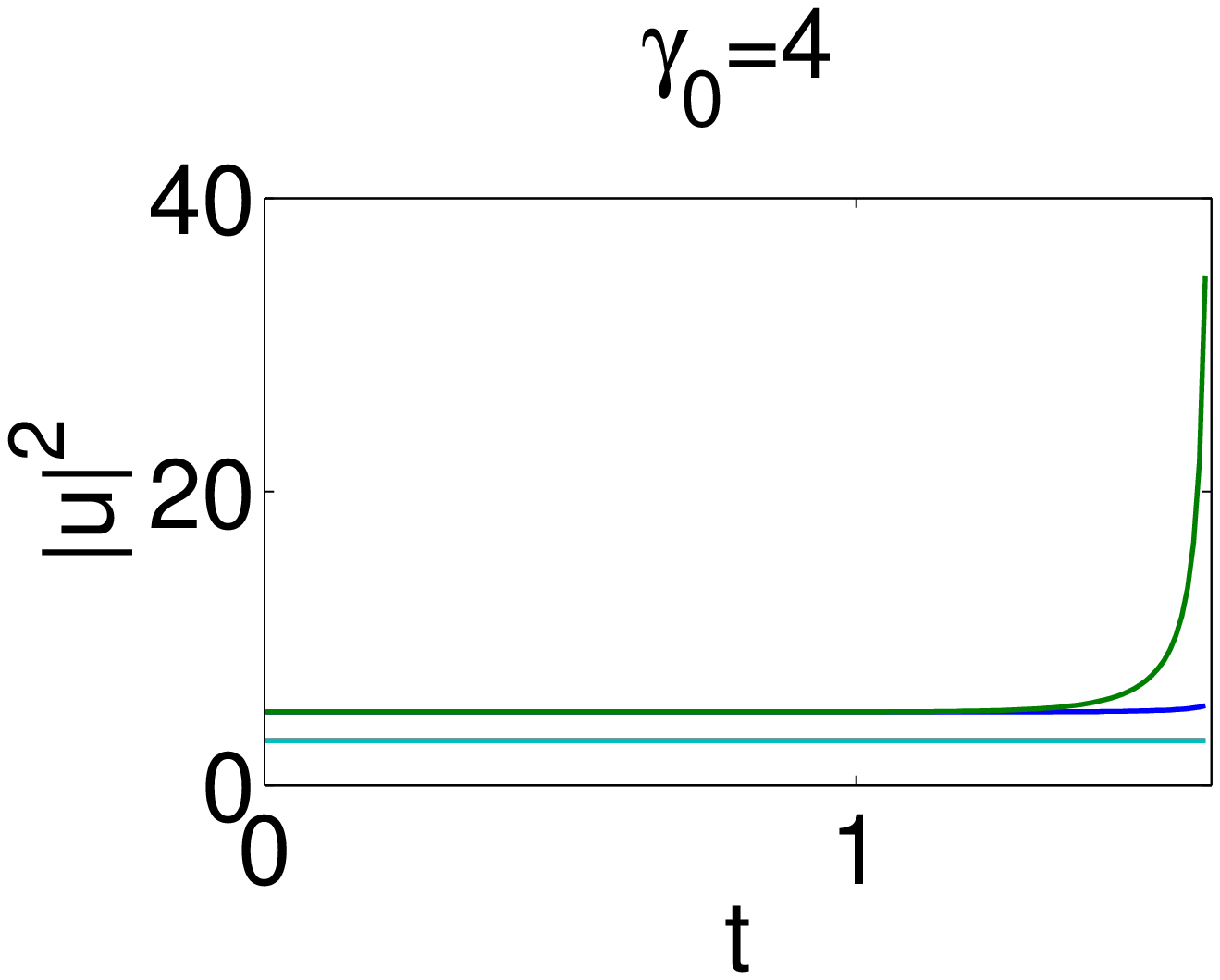}}}
\subfigure[\ thick red
branch]{\scalebox{0.4}{\includegraphics{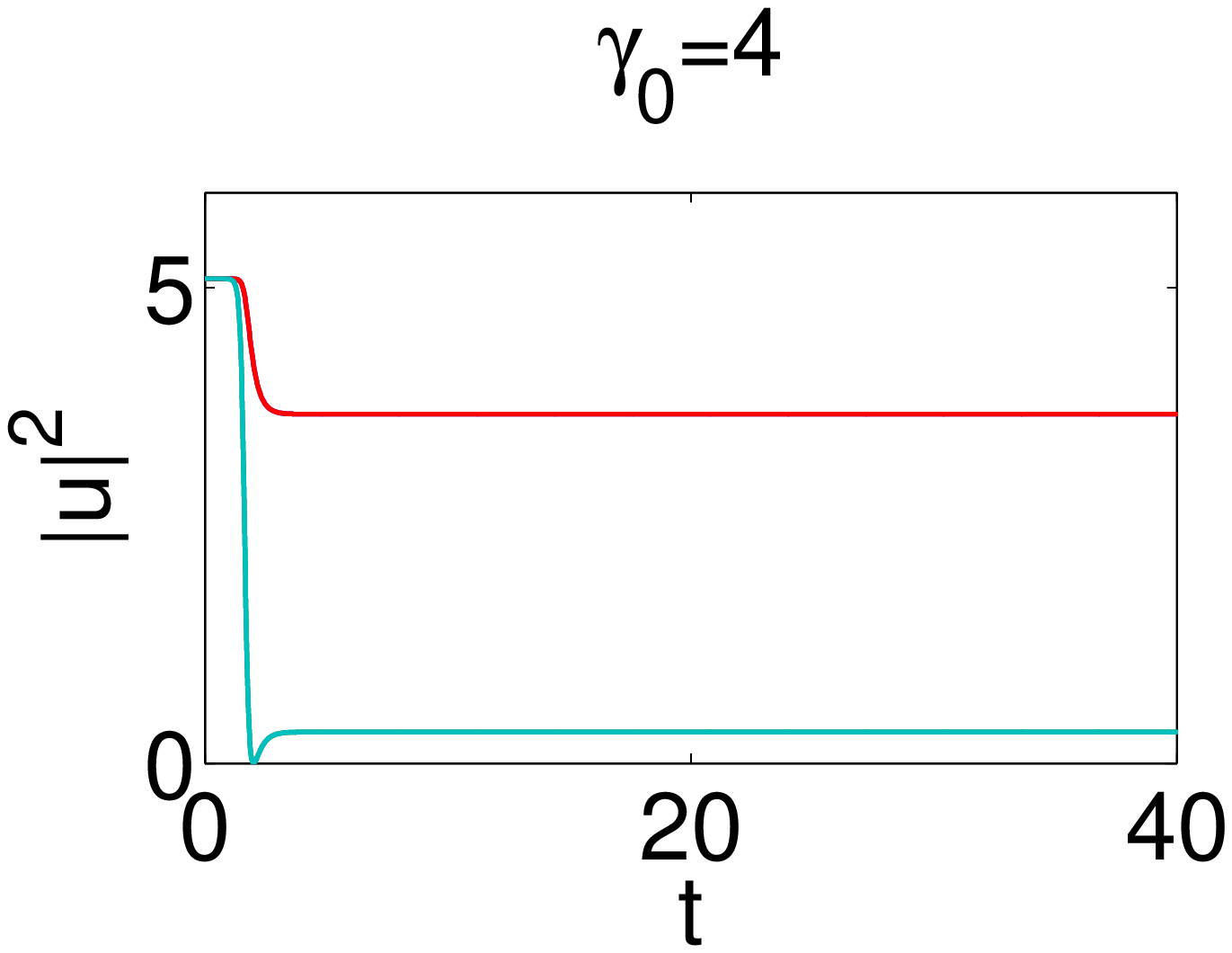}}}
\caption{(Color online) Perturbed evolution of the solutions for the
cross-coupled bi-dimers corresponding to Fig.~\protect\ref{cross}.}
\label{cross_dyn}
\end{figure}

\section{Conclusion}

We have introduced two settings bearing both linear and nonlinear gain and
loss, which implement, in the simplest form, different fundamental\ types of
symmetries for $\mathcal{PT}$-invariant systems. Both settings are built of
two linearly coupled intrinsically nonlinear $\mathcal{PT}$-symmetric
dipoles (which, by themselves, provide for the simplest implementations of
the $\mathcal{PT}$ invariance). One of the configurations is arranged as a
straight-coupled bi-dimer, with each linear-gain site linearly-coupled to
another linear-gain one, and a linear-loss site. The second configuration is
the cross-coupled bi-dimer, with each of the gain sites coupled to two lossy
ones. The latter system may also implement a $\mathcal{PT}$-hypersymmetric
quadrimer, in the case when all the linear-coupling coefficients are equal.

For these two systems, we have identified a number of solutions
analytically, including those keeping both the spatial and the $\mathcal{PT}$%
-symmetries unbroken, as well as solutions that break one of these
symmetries. These solutions present a number of noteworthy features, in
terms of the bifurcation theory. In particular, symmetry-breaking and
saddle-center bifurcations were found in these settings. Instabilities
typically lead to blowup of the solutions, but examples of convergence to
another attractor were also identified.

These fundamental systems may be used as building blocks to construct
lattices consisting of either straight- or cross-coupled bi-dimers, which
should be a natural next step of the analysis. It would also be of
particular interest to extend considerations to a three-dimensional $%
\mathcal{PT}$-symmetric system, i.e., to explore $\mathcal{PT}$-symmetric
\textit{cubes} and configurations which may be developed from them. Studies
along these directions are currently in progress.

\vspace{5mm}

\acknowledgments PGK gratefully acknowledges support from the US National
Science Foundation under grants CMMI-1000337 and DMS-1312856, and from the
US-AFOSR under grant FA9550-12-1-0332. PGK and BAM appreciate a partial
support provided by the Binational (US-Israel) Science Foundation through
grant 2010239.

\vspace{2cm}

\begin{center}
\uppercase{University of Massachusetts \\
Tel Aviv University} \\
\vspace{5mm}
(Received Sep 26, 2013)
\end{center}
\end{document}